\documentclass{emulateapj}

\usepackage{amssymb}
\usepackage{amsmath}
\usepackage{epsfig}
\usepackage{graphicx}
\usepackage{natbib}

\slugcomment{accepted by ApJ Letters June 23, 2009}

\begin{document}

\title{
Collisional Stripping and Disruption of Super-Earths
}
\author{Robert A. Marcus\altaffilmark{1,A}, Sarah T. Stewart\altaffilmark{2},
Dimitar Sasselov\altaffilmark{1}, Lars Hernquist\altaffilmark{1}
}
\affil{$^1$Astronomy Department, Harvard University, Cambridge, MA 02138}
\affil{$^2$Department of Earth and Planetary Sciences, Harvard University, 
Cambridge, MA 02138}
\email{$^a$rmarcus@cfa.harvard.edu}

\begin{abstract}
  The final stage of planet formation is dominated by collisions
  between planetary embryos. The dynamics of this stage determine the
  orbital configuration and the mass and composition of planets in the
  system. In the solar system, late giant impacts have been proposed
  for Mercury, Earth, Mars, and Pluto. In the case of Mercury, this
  giant impact may have significantly altered the bulk composition of
  the planet.  Here we present the results of smoothed particle
  hydrodynamics simulations of high-velocity (up to $\sim 5 v_{esc}$)
  collisions between 1 and 10 $M_{\oplus}$ planets of initially
  terrestrial composition to investigate the end stages of formation
  of extrasolar super-Earths.  As found in previous simulations of
  collisions between smaller bodies, when collision energies exceed
  simple merging, giant impacts are divided into two regimes: (1)
  disruption and (2) hit-and-run (a grazing inelastic collision and
  projectile escape). Disruption occurs when the impact parameter is
  near zero, when the projectile mass is small compared to the target,
  or at extremely high velocities. In the disruption regime, we derive
  the criteria for catastrophic disruption (when half the total
  colliding mass is lost), the transition energy between accretion and
  erosion, and a scaling law for the change in bulk composition
  (iron-to-silicate ratio) resulting from collisional stripping of a
  mantle.
\end{abstract}


\section{Introduction}

To date, more than 300 extrasolar planets have been discovered.  Of
these, more than 10 have masses $\lesssim 10 M_{\oplus}$.  With the
launch of the COROT \citep{COROT:2003} and Kepler \citep{Kepler:2003}
satellites, it is expected that many transiting super-Earths will be
discovered in the next few years. In fact, the first transiting
super-Earth candidate, CoRoT-7b, has been announced (Leger et al, in
prep). The radius determined from the transit, when combined with
radial velocity measurements of the planet's mass, can be used to
determine a mean density and thus infer a bulk composition
\citep{Valencia:2007b}.

Although the chemical composition of planets in the solar system
generally reflects the gradient in the nebula, the final collisions
forming each planet may involve embryos scattered from regions with
different bulk composition; hence the final composition of each planet
is thought to be dominated by the last few impact events
\citep[e.g.,][]{Wetherill:1994,Chambers:2004}.  Planet formation
simulations have found that late stage impacts can reach velocities up
to six times the mutual escape velocity in the terrestrial planet
region of the solar system \citep{Agnor:1999}.  Due to the presence of
giant planets and planetary migration, impact velocities may be even
higher in other planetary systems \citep[e.g.,][]{Raymond:2005a,
  Mandell:2007}. Thus, the formation of super-Earths may involve very
high velocity impacts between large bodies (e.g., 0.1 to several Earth
masses). Collision outcomes will vary widely depending on the impact
velocity, mass ratio between the bodies, and the impact angle
\citep{Agnor_Asphaug:2004, Asphaug:2009, Stewart_Leinhardt:2009}. The
conditions for imperfect merging are particularly interesting as they
provide an opportunity to further alter the bulk compositions of the
growing planets. An erosive impact event may be responsible for the
high bulk density of Mercury relative to the other terrestrial planets
in the solar system \citep{Benz:1988, Benz:2007}. In addition,
collisional erosion has been considered for the Earth-Moon impact
\citep{ONeill_Palme:2008}.

The end stages of planet formation are typically studied using $N$-body
codes that assume perfect merging of the masses of colliding embryos
\citep[e.g.,][]{Agnor:1999,Chambers:2009}. However, \citet{Agnor:1999}
showed that the assumption of perfect merging often leads to the
combined body rotating faster than the critical limit.  Recently,
various hybrid techniques combine $N$-body dynamics and a statistical
treatment of fragmentation \citep[e.g.,][]{Charnoz_Morbidelli:2003,
  Bromley_Kenyon:2006}.  The adopted fragmentation models are usually
derived from simulations of erosive collisions over a range of target
sizes and impact velocities \citep[e.g.,][]{Benz_Asphaug:1999,
  Benz:2000, Stewart_Leinhardt:2009}.  Models of planetary accretion
(still assuming perfect merging) now encompass the formation of
extrasolar systems with super-Earths and giant planets
\citep[e.g.,][]{Thommes:2008}. Hence, there is a need for simple
descriptions of the outcome of collisions between large planetary
embryos.

To complicate matters, recent studies of collisions between planetary
embryos describe two dramatically different regimes when the impact
energies exceed perfect merging \citep{Agnor_Asphaug:2004,
  Asphaug:2006, Asphaug:2009}: (1) erosion via disruption and
gravitational dispersal and (2) ``hit-and-run'' events where the
projectile grazes the target, which is left nearly intact. The
boundary between the two regimes depends on the mass ratio and impact
angle \citep[see Figure 17 in][]{Asphaug:2009}. The disruption regime
is typically characterized by the catastrophic disruption criteria,
$Q^{*}$, which is a size-dependent curve describing the projectile
kinetic energy divided by the target mass such that half the target
mass remains. For small bodies ($\lesssim$~1~km in diameter), tensile
strength, which decreases with size, determines the outcome. For
larger bodies, gravity dominates and the largest remnant includes
reaccumulated material \citep{Melosh_Ryan:1997}.  The transition
between accretion (net growth of the target) and erosion is determined
by the dependence of the mass of the largest remnant on the impact
energy.  \citet{Benz_Asphaug:1999} simulated colliding rocky and icy
bodies to determine the catastrophic disruption criteria from scales
of centimeters to hundreds of kilometers. They found that the ratio of
the largest remnant mass, $M_{lr}$, to the target mass, $M_{targ}$,
depends linearly on the impact energy divided by target mass, $Q$,
scaled to $Q^{*}$.  This simple relationship has been confounded by
the fact that $Q^{*}$ depends on the target size, impact velocity,
mass ratio, and material properties.  Thus, previous work relied upon
the availability of catastrophic disruption criteria for specific
impact scenarios. However, \citet{Stewart_Leinhardt:2009} have
recently developed a more general expression for the disruption
criteria.  The ratio $M_{lr}/M_{targ}$ is replaced by
$M_{lr}/M_{tot}$, where $M_{tot}$ is the total colliding mass. The
impact energy divided by target mass, $Q$, is replaced by the reduced
mass kinetic energy scaled to the total colliding mass, $Q_{R} =
\frac{1}{2}\mu V_{i}^{2}/M_{tot}$, where $\mu$ is the reduced mass and
$V_{i}$ is the impact velocity. Using these variables, the new
catastrophic disruption criteria, $Q^{*}_{RD}$, at which
$M_{lr}/M_{tot}=0.5$ is not dependent on the mass ratio. Additionally,
\citet{Stewart_Leinhardt:2009} provide parameters that describe a wide
range of material properties and impact velocities. However, to date,
the catastrophic disruption criteria has not been derived for bodies
larger than 100 km.

Here, we simulate a wide range of impacts between planetary embryos up
to and including super-Earths to derive the conditions for accretion
versus erosion.  We consider bodies with initial compositions similar
to Earth and derive scaling laws for catastrophic disruption and
changes to the iron-to-silicate mass ratio resulting from erosion.

\section{Method}

\subsection{Smoothed Particle Hydrodynamics (SPH)}

Hydrodynamic codes have been used extensively to simulate giant
impacts. SPH is a Lagrangian technique in which the mass distribution
is represented by a finite set of particles evolved with time
\citep{Gingold_Monaghan:1977, Lucy:1977}.  To date, SPH has been used
to study giant impacts in the early histories of Earth
\citep{Benz:1986, Canup:2004}, Mercury \citep{Benz:1988, Benz:2007},
Mars \citep{Marinova:2008}, and Pluto \citep{Canup:2005}. These
studies have all utilized an SPH code descended from that of
\citet{Benz:1986}.

Here we present the results of simulations of giant impacts using the
SPH code GADGET \citep{Springel:2005}, which has been tested and
utilized extensively in cosmology and galactic dynamics.  In
particular, GADGET employs a tree-based scheme for computing the
self-gravity of objects \citep{Hernquist:1989}, and, unlike other
implementations of SPH, a fully conservative approach for integrating
the hydrodynamic equations \citep{Springel:2002} that maintains energy
and entropy conservation even when smoothing lengths vary adaptively
\citep{Hernquist:1993, OShea:2005}.

We have modified GADGET to compute thermodynamic quantities by
interpolating between elements of tabulated equations of state. The
tables were generated using a revised version of the semianalytic
equation of state model ANEOS \citep{Thompson_Lauson:1972,
  Melosh:2007}. We used ANEOS parameters for SiO$_{2}$
\citep{Melosh:2007}, forsterite, and iron (Melosh, private
communication). The EOS tables were finely gridded to resolve phase
boundaries \citep[in a manner similar
to][appendix]{Senft_Stewart:2008}. The modified code was tested with
three-dimensional simulations of impacts between plates of identical
material at multiple resolutions and shock pressures. The calculated
peak pressures agree with the analytic impedance match solution
\citep[][ p.~54]{Melosh:1989}. We performed two additional tests of
the code, which are described below. As in previous studies of giant
impacts \citep[e.g.,][]{Benz:1986}, shear and tensile strength are
neglected because gravity is expected to dominate in this size regime.

\subsection{Test One: Shock Pressure Decay}

When a body impacts a half space, beyond an initial isobaric region
near the impact point, the peak shock pressure decays exponentially
with distance, $d$, via
\begin{equation} 
P \propto d^{-n}, \label{eq:decay}
\end{equation}
where $n$ is a function of impact velocity.  Using a two-dimensional
Eulerian hydrocode and a range of materials, \citet{Pierazzo:1997}
found that $n = (-1.84 \pm 0.17) + (2.61 \pm 0.14)$ log($V_{i}$),
where $V_{i}$ is the impact velocity in km s$^{-1}$.

\begin{figure}
\figurenum{1}
\begin{center}
\includegraphics[scale=0.5]{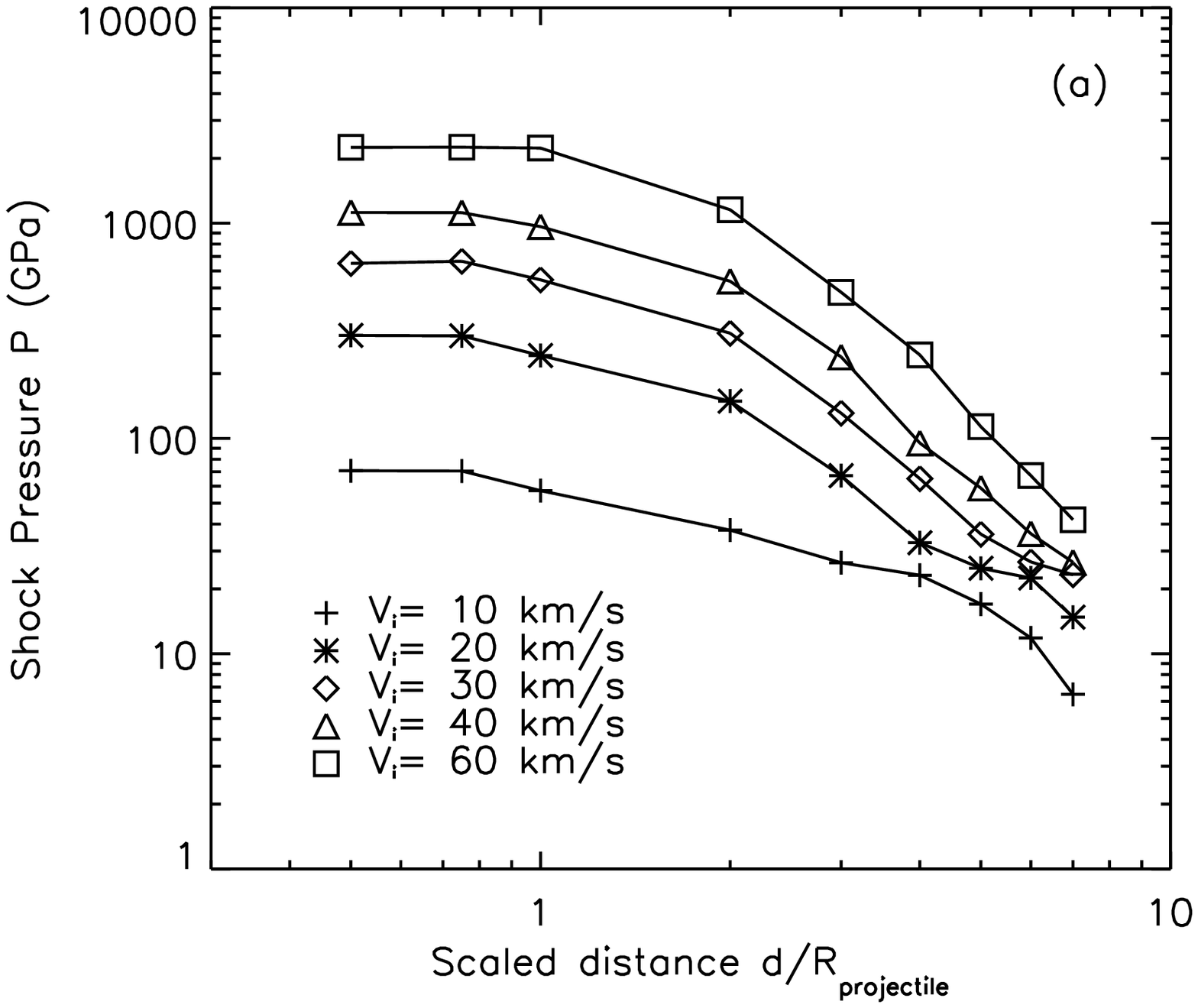}
\includegraphics[scale=0.5]{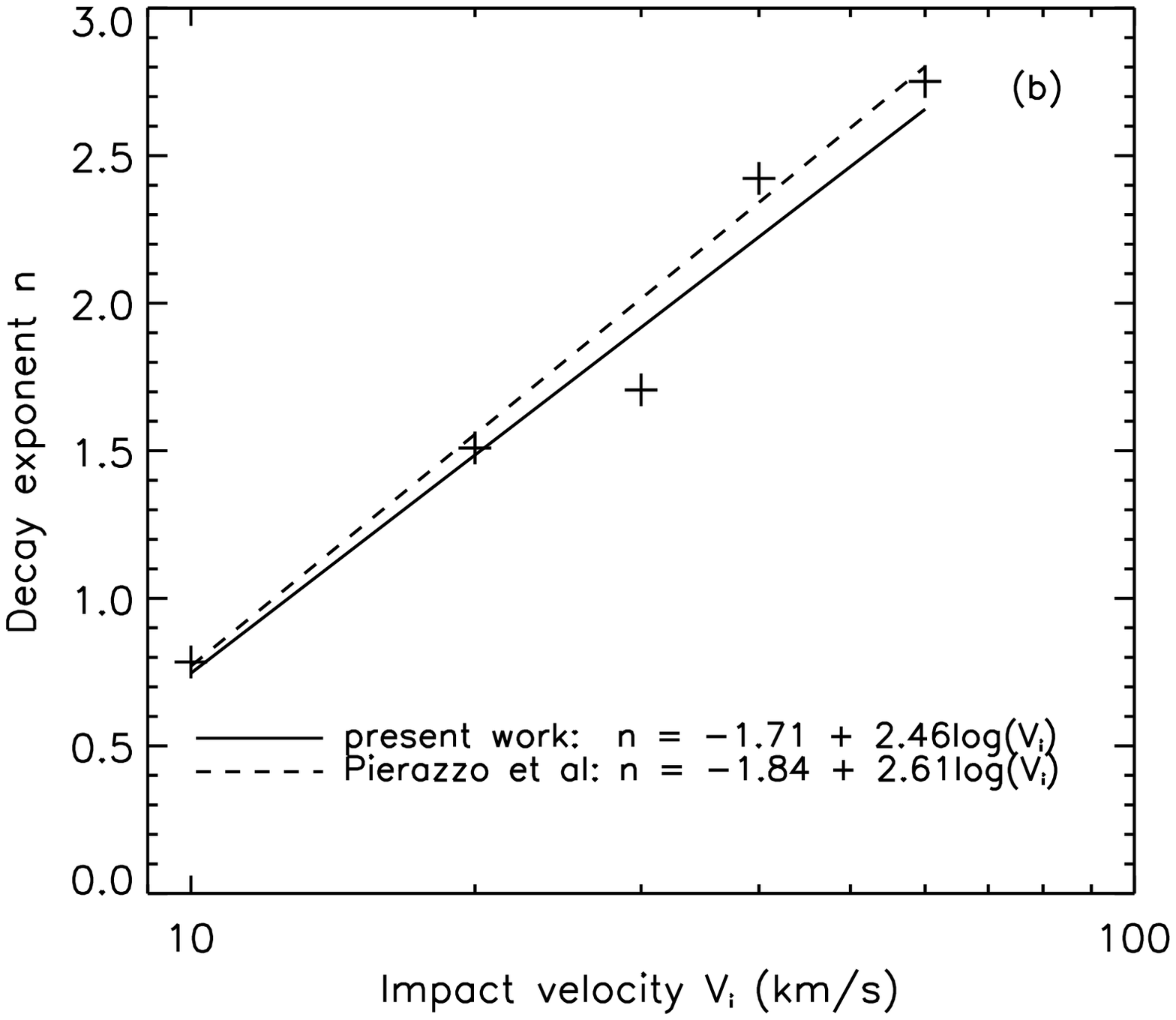}
\end{center}
\caption{Test One: Shock Pressure Decay. (a) Shock pressure as a
  function of scaled distance from impact point for a SiO$_{2}$ target
  and projectile.  (b) Power-law decay exponent
  (equation~\ref{eq:decay}) as a function of impact velocity, with
  fits from present work and \cite{Pierazzo:1997}. }
\label{ref:decay}
\end{figure}

Using our modified version of GADGET, we reproduced the simulations of
\citet{Pierazzo:1997} for SiO$_{2}$ and iron. We modeled the pressure
decay profile from a $10$ km diameter impactor striking a large flat
surface (essentially a half space) at speeds of $10$, $20$, $30$,
$40$, and $60$ km s$^{-1}$, shown in Figure~\ref{ref:decay}a.  We fit
the exponent $n$ with $(-1.70 \pm 0.33) + (2.45 \pm
0.24)$log($V_{i}$), in excellent agreement with previous work
(Figure~\ref{ref:decay}b).

\subsection{Test Two: Moon-Forming Impact}

The iron depletion in the Moon, isotopic similarities between Earth
and Moon, and the large angular momentum of the Earth-Moon system led
to the giant impact hypothesis for the origin of the Moon
\citep{Hartmann_Davis:1975, Cameron_Ward:1976}.  SPH has been
extensively employed in modeling potential Lunar-forming impacts, most
recently by \citet{Canup:2004, Canup:2008}.

We repeated two of the simulations from Table 1 of \citet{Canup:2004}.
We started from the same initial conditions: a target and impactor of
70\% silicates (forsterite) and 30\% iron, hydrostatic pressure
profile, and an initial isentropic temperature profile with a surface
temperature of 2000~K. The target and impactor were allowed to settle
to negligible particle velocities in isolation. The final masses of
the proto-Earth and proto-Lunar disk were calculated using the
iterative procedure described in the appendix of \citet{Canup:2001}.
Table \ref{ref:MFI} presents our results from these two simulations,
which are in excellent agreement with those of \citet{Canup:2004}.

\subsection{Present Work}

We performed a series of more than 60 simulations of impacts onto
super-Earth sized bodies. The initial conditions were targets and
impactors of roughly terrestrial composition, comprised of 67\%
forsterite and 33\% iron. The bodies were initialized with hydrostatic
pressure profiles and temperature profiles calculated for super-Earths
with surface temperatures of about 350~K \citep[][Figure
4]{Valencia:2006}. Previous work did not show a strong dependence on
initial temperature \citep{Benz:2007}.
The bodies were allowed to settle to negligible particle velocities in
isolation with the temperature fixed.  We used three different target
masses, 1$M_{\oplus}$, 5$M_{\oplus}$, and 10$M_{\oplus}$, with
impactors of 1/4, 1/2, and 3/4 the target mass.  All the initial
experiments employed an impact parameter of zero, and a subset was
repeated with impact angles of $45^{\circ}$. A few cases of
undifferentiated, pure forsterite bodies were considered as a more
direct comparison to previous work on catastrophic disruption, which
has not included differentiated bodies.  The number of particles
ranged from $1.25 \times 10^{5}$ to $7 \times 10^{5}$ (30-45 particles
per target diameter), and the results were checked for sensitivity to
resolution.

The mass of the largest remnant following each impact was determined
using a fragment search algorithm similar to that employed by
\citet{Benz_Asphaug:1999}.  The potential and kinetic energies of all
particles were calculated with respect to the particle closest to the
potential minimum. The center of mass position and velocity were then
calculated for all bound particles and the process was repeated on the
remaining unbound particles until the calculation converged (typically
after four or five iterations).

\section{Results}

Our results fall into two regimes: (1) disruptive events and (2)
hit-and-run events.  At impact energies just above merging for events
with projectile to target mass ratios ranging from 1/10 to 1, the
results from \citet{Asphaug:2009} indicate that the onset of
hit-and-run events occurs at impact angles between 30$^{\circ}$ and
45$^{\circ}$ (where $0^{\circ}$ is head-on). Our results are in
agreement and all the 45$^{\circ}$ impact cases were merging or
hit-and-run events.

\subsection{Catastrophic Disruption: Scaling largest remnant and iron composition}

\begin{figure} 
\figurenum{2}
\begin{center}
\includegraphics[scale=0.6]{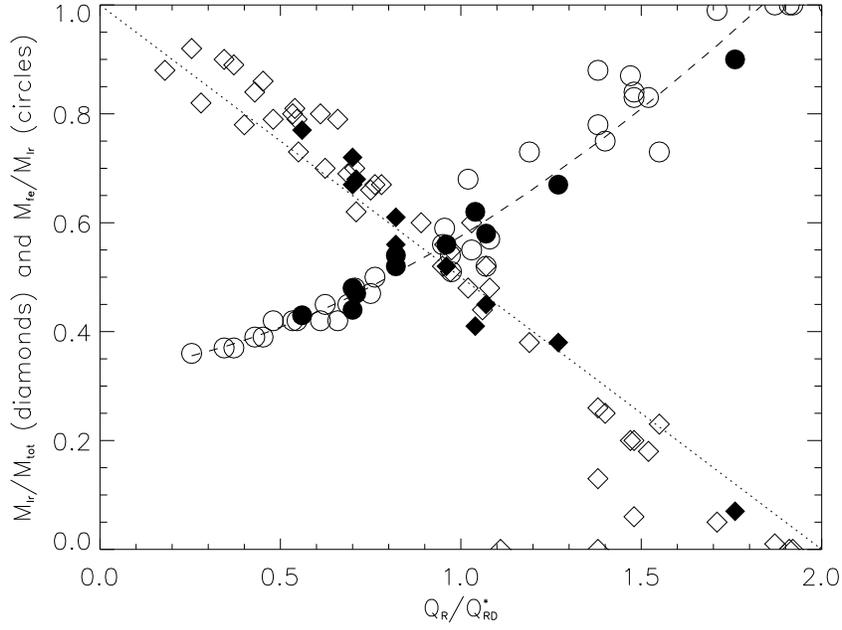}
\end{center}
\caption{Largest remnant mass and iron fraction vs.\ scaled impact
  energy.  Open symbols, this work; filled symbols, \cite{Benz:1988,
    Benz:2007}. Dotted line, Equation~\ref{ref:mlr}; dashed curve,
  Equation~\ref{ref:fe_fraction}.}
\label{ref:mlr}
\end{figure}

In the disruption regime, the outcome of a collision is described in
terms of the largest remnant mass and the catastrophic
disruption criteria.  In Figure \ref{ref:mlr}, we present the largest
remnant mass as a function of the scaled impact energy (diamonds) for
our head-on impact events (open symbols) and a subset of simulations
from candidate Mercury forming impacts from
\citet{Benz:1988,Benz:2007} (filled symbols). The results are in
excellent agreement with the universal law for the largest remnant
mass presented by \citet{Stewart_Leinhardt:2009} (dotted line), where
\begin{equation}
M_{lr}/M_{tot} = -0.5(Q_{R}/Q^{*}_{RD} -1) + 0.5 \, .
\label{ref:mlreq}
\end{equation}
This linear relationship derived by \citet{Stewart_Leinhardt:2009} for
smaller bodies and slower impact velocities continues to hold into the
super-Earth regime. Note that the utility of equation~\ref{ref:mlreq}
is bounded by $Q_{R}/Q^{*}_{RD}$ of 0 (events near merging) and 2 (a
super-catastrophic regime).

We find that the iron mass fraction of the largest remnant follows a
simple power law (circles in Figure~\ref{ref:mlr}), given by
\begin{equation} 
M_{Fe}/M_{lr} = 0.33 + 0.25(Q_{R}/Q^{*}_{RD})^{1.65}.
\label{ref:fe_fraction}
\end{equation} 
Note that we included only the subset of simulations from Benz from
which the catastrophic disruption criteria could be derived.

\begin{figure}
\figurenum{3}
\begin{center}
\includegraphics[scale=0.6]{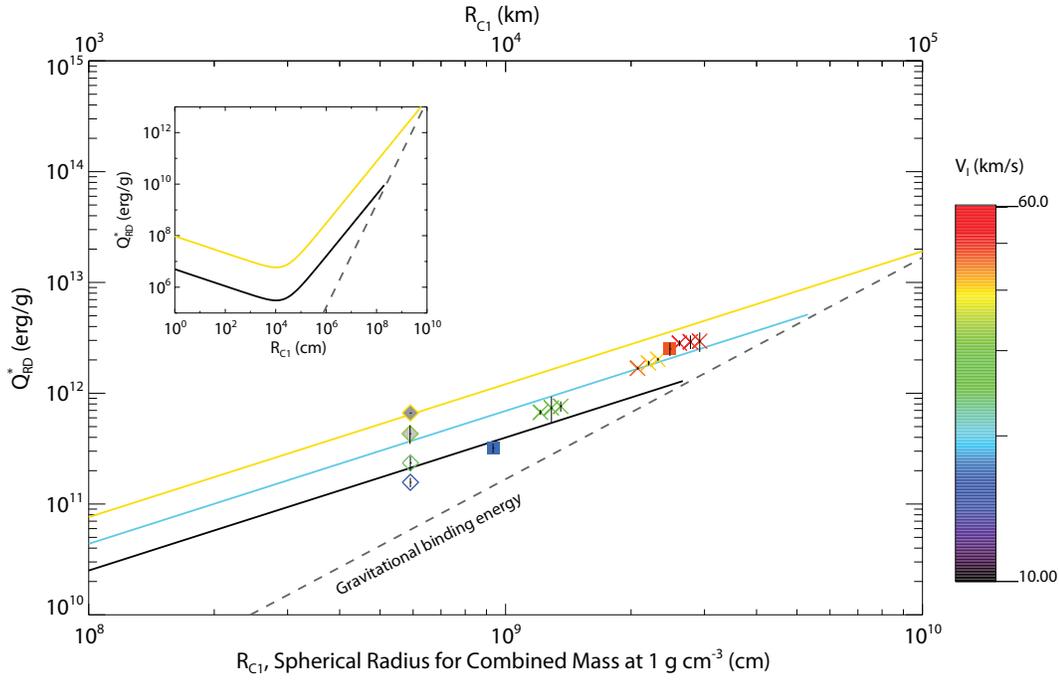}
\end{center}
\caption{Catastrophic disruption criteria $Q^{*}_{RD}$ vs.\ $RC_1$.
  Results from impacts between differentiated bodies ($\times$) and
  pure forsterite bodies (squares).  Diamonds are Mercury-size events
  from \citet{Benz:1988, Benz:2007} where shading indicates
  0$^{\circ}$, 30$^{\circ}$ and 45$^{\circ}$ impacts. Symbol colors
  indicate impact velocity. Vertical black bars are 1$\sigma$ errors
  on $Q^{*}_{RD}$. Solid lines are disruption criteria for impact
  velocities of 10 (black), 20 (blue), and 40 km s$^{-1}$ (orange)
  (Equation~\ref{eq:qsrd}). The inset shows full size range for the
  velocity-dependent catastrophic disruption curves from
  \cite{Stewart_Leinhardt:2009}. }
\label{ref:q*rd}
\end{figure}

The fitted catastrophic disruption criteria from our simulations and
the subset from \citet{Benz:1988, Benz:2007} are shown in
Figure~\ref{ref:q*rd}.  The lines are the gravity-regime
velocity-dependent catastrophic disruption criteria for low strength
bodies from \cite{Stewart_Leinhardt:2009} (in cgs units):
\begin{equation}
Q^{*}_{RD} = 10^{-4} R_{C1}^{1.2} V_{i}^{0.8}. 
\label{eq:qsrd}
\end{equation}
$R_{C1}$ is the radius of a spherical body with $M_{tot}$ and density
of 1 g cm$^{-3}$. The $Q^{*}_{RD}$ data points for giant impacts fall
short of the prediction from Equation~\ref{eq:qsrd} by less than a
factor of 2. Hence, the results for super-Earths are entirely
consistent with disruption criteria developed for smaller bodies in
the gravity regime, and one can reasonably determine the largest
remnant mass and bulk composition following a collision knowing only
the total colliding mass and impact velocity.

\subsection{Disruption Versus hit-and-run impacts}

Our high velocity impacts at angles of $45^{\circ}$ by projectile
masses between 1/4 and 3/4 the target mass produced no significant
change to the mass or the bulk composition of the target.  In such
cases, there was only an inelastic collision resulting in the target
heating and acquiring angular momentum. Hence, we independently verify
the phenomenon described as hit-and-run by \citet{Asphaug:2006}.

\begin{figure}
\figurenum{4}
\begin{center}
\includegraphics[scale=0.4]{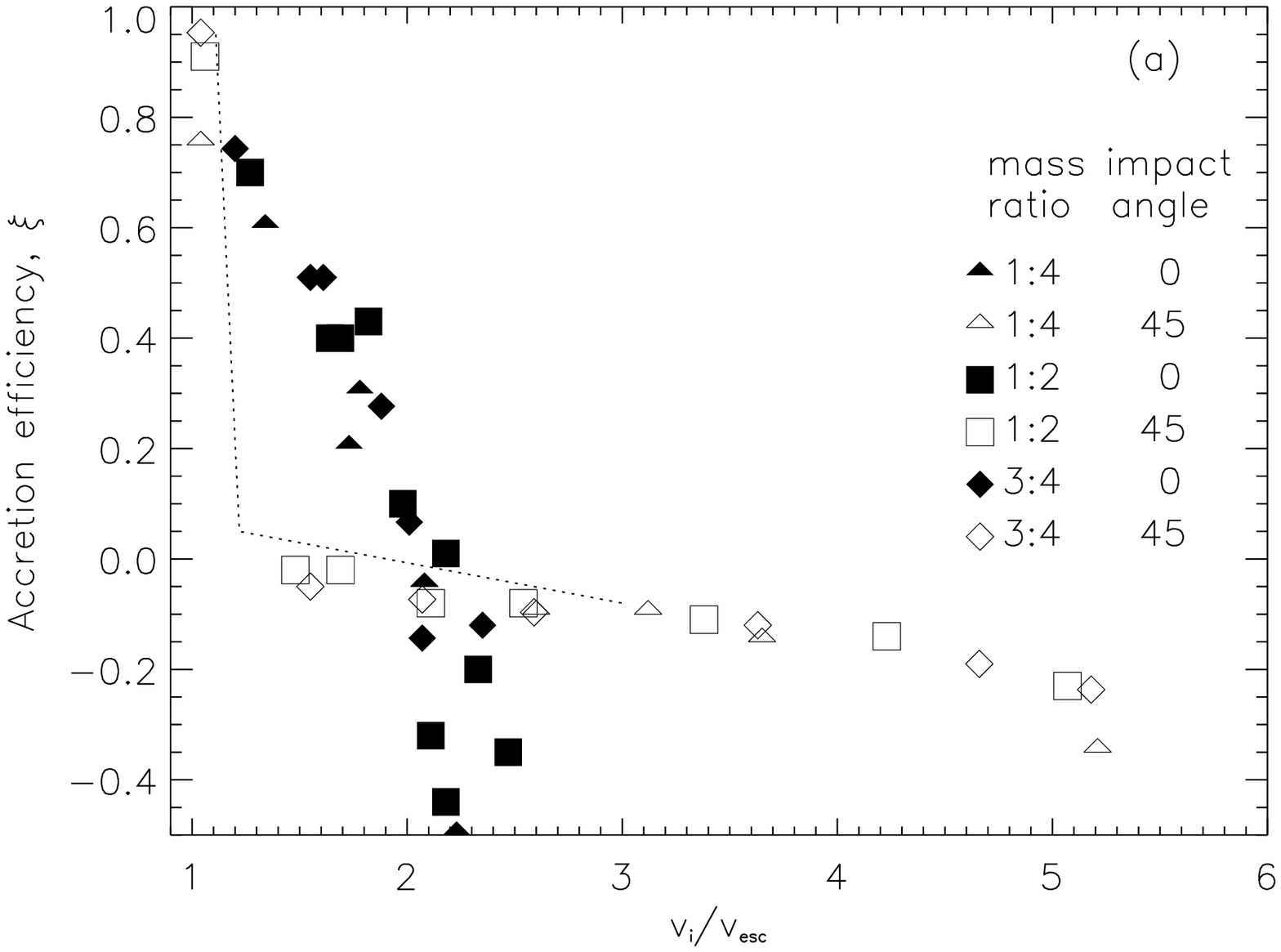}
\includegraphics[scale=0.4]{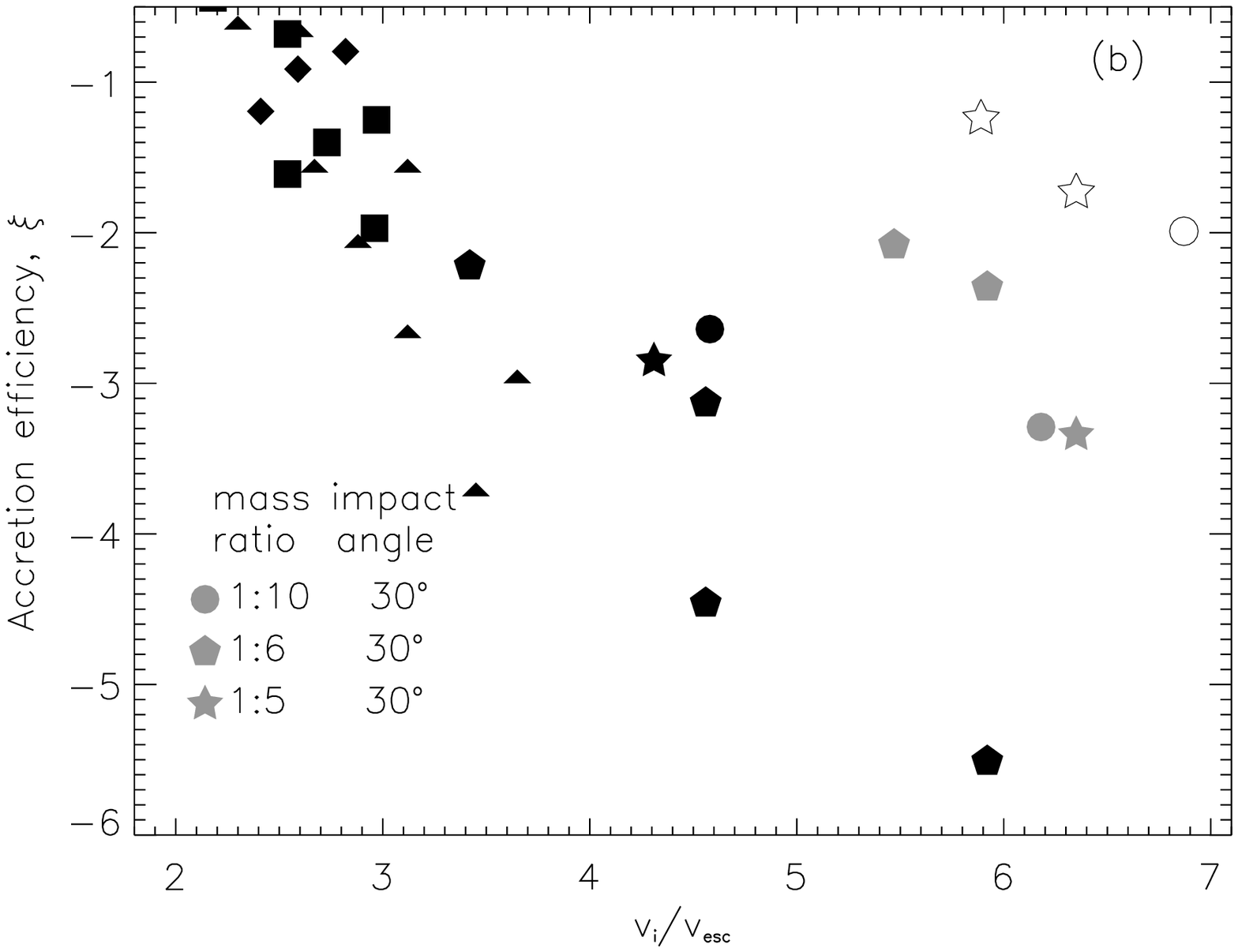}
\end{center}
\caption{Accretion efficiency, $\xi$, as a function of scaled impact
  velocity. (a) Accretion regime. Collision outcome depends strongly
  on impact angle. Dashed line presents results for $45^{\circ}$, 1:2
  mass ratio impacts from \citet[][Figure 17]{Asphaug:2009}.  (b)
  Disruption regime. Catastrophic disruption is $\xi=-4.5$, -2.0, and
  -0.5 for 1:10, 1:5, and 1:2 mass ratios, respectively. In both
  panels, symbols denote mass ratio and shading denotes impact angle.
  Circles, pentagons, stars -- from \cite{Benz:1988,Benz:2007}. Other
  symbols -- this work.  }
\label{ref:vesc}
\end{figure}

The contrast between the disruption and hit-and-run regimes is
illustrated by the accretion efficiency defined by
\citet{Asphaug:2009},
\begin{equation}
\xi \equiv \frac{M_{lr} - M_{targ}}{M_{proj}},
\label{ref:accretion}
\end{equation}
where positive values correspond to growth and negative values to
erosion of the target. Note that the value of $\xi$ that corresponds
to catastrophic disruption depends on the projectile to target mass
ratio ($-0.5M_{targ}/M_{proj}+0.5$). Figure \ref{ref:vesc} presents
accretion efficiency as a function of impact velocity, scaled to the
mutual escape velocity ($v_{esc}^{2} =
2G(M_{targ}+M_{proj})/(R_{targ}+R_{proj})$).  For head-on
impacts (filled black symbols), the fraction of the colliding mass
remaining in the largest remnant decreases steadily with increasing
$v_i/v_{esc}$. However, for $45^{\circ}$ impacts (open symbols), the
outcome is nearly a step function from perfect merging to remnant
masses approximately equal to the target mass ($\xi = 0$).
Our results for super-Earths are in excellent agreement with the
studies of accretion efficiency for impacts onto much smaller bodies
(0.10$M_{\oplus}$) presented in \citet{Agnor_Asphaug:2004} and
\citet{Asphaug:2009} (e.g., dotted line in Figure~\ref{ref:vesc}a).
Note that the hit-and-run regime encompasses a very large range of
impact velocities.  In this work, we considered events exceeding
$5v_{esc}$, about twice as fast as in previous work.  Hence, the
possible outcomes of impacts between comparably sized bodies are
similar over orders of magnitude in target mass.

At sufficiently high impact energies, the hit-and-run regime
transitions to disruption.  A study of collisions onto 100 km
diameter bodies by projectile masses up to a ratio of 1:10 found that
the impact energies required for catastrophic disruption may increase
by an order of magnitude or more with increasing impact angle
\citep[Figure 1 in][]{Durda:2007}.
The impact angle dependence is seen in Figure~\ref{ref:vesc}b: near
the catastrophic disruption value of $\xi=-2$ for 1:5 mass ratio
(stars), compare the offset in velocity for impact angles of
0$^{\circ}$, 30$^{\circ}$, and 45$^{\circ}$.  The increased disruption
energy as a function of impact angle is given in the diamond points in
Figure~\ref{ref:q*rd}.

Thus, the results of giant impact events at velocities up to several
times $v_{esc}$ depend primarily on (1) the impact angle and mass
ratio which define the transition between the disruption and
hit-and-run regimes and (2) the total colliding mass and impact
velocity which define $Q^*_{RD}$ in the disruption regime. The outcome
falls in the disruption regime at any impact angle when the projectile
is very small and fast but only when the impact angle is small
(typically $\lesssim30^{\circ}$) when the projectile and target are
comparably sized. Because $45^{\circ}$ is the most probable impact
angle, catastrophic disruption of protoplanets is most likely to occur
when the impactor is much smaller than the target.

\section{Discussion and Conclusions}

Using the results from this work and previous studies of giant
impacts, we have described the outcome of collisions between a wide
range of masses, from planetary embryos to super-Earths.  In the
disruptive regime, the framework developed by
\citet{Stewart_Leinhardt:2009} to describe the transition between
accretion versus erosion in hypervelocity impacts holds from the lab
scale, with sizes in centimeters and velocities in m s$^{-1}$, up to
the scales of the largest expected super-Earths, with sizes up to ten
thousand kilometers.

In the disruption regime, we derived a scaling law for the changes in
bulk composition resulting from collisions between differentiated
bodies of Earth-like composition. Based on the boundary between
disruptive and hit-and-run type events, significant compositional
changes require either a small impact angle or a small, fast
projectile. Because of this narrow range of impact conditions that can
lead to an increase in bulk density, the presence of unusually dense
planets can place constraints on the dynamical history of a planetary
system.  In particular, if planet growth occurs by merging
similar-sized bodies, changes to the bulk composition of the target
are unlikely.  Thus, one-time high-energy impacts provide the best
means of increasing a planet's bulk density. If the specifics of the
planetary system rule out such a collision,
then such a planet is more likely to have formed in a protoplanetary
disk with substantially different chemical abundances or with a
density profile quite different from the minimum-mass solar
nebula \citep{Raymond:2005b}.

For very energetic impacts, the largest remnant can have a
substantially different composition than that of Earth, even when the
target and impactor initially had Earth-like bulk compositions.  This
is consistent with the hypothesis that Mercury, which is $\sim 70\%$
iron by mass, owes its present bulk composition to a catastrophic
impact event. For planet masses close to that of Mercury, the critical
velocity needed to yield a remnant with such an iron enrichment is
about 10-20 km s$^{-1}$, as was found by \citet{Benz:1988, Benz:2007}.
Such velocities are smaller than the orbital velocities inside 1 AU
for a $\sim 1 M_{\odot}$ star.  However, for more massive planets, the
critical velocities can enter the 50-60 km s$^{-1}$ range, requiring
extreme conditions for a Mercury-like planet to be created.
Thus, Mercury-like planets should be much more common than
super-Mercuries, although even these may form in very close orbits in
extrasolar systems.  Hence, the probability of forming super-Mercuries
warrants further investigation (Marcus et al., in prep).

We thank Joe Barranco for help in code development and Zoe Leinhardt,
T.J. Cox and the anonymous reviewer for useful feedback. The
simulations in this Letter were run on the Harvard FAS Odyssey
cluster.

\pagebreak

\begin{deluxetable}{cccccccccc}
  \tablewidth{0pt} \tablecolumns{10} \tablecaption{Test Two: Potential
    Moon forming impacts.}

  \tablehead{ \colhead{$N/10^{3}$} &
    \colhead{$\gamma$} & \colhead{$v$} & \colhead{b} &
    \colhead{$M_D/M_L$} & \colhead{$L_D/L_{EM}$} &
    \colhead{$M_{Fe}/M_D$} & \colhead{$M_{imp}/M_D$} &
    \colhead{$L_F/L_{EM}$} & \colhead{$M_M/M_L$}
  } 
  \startdata
  \multicolumn{10}{c}{This work} \\
  120 & 0.13 & 1.00 & 0.730 & 1.40 & 0.27 & 0.04 & 0.78 & 1.06 & 1.24 \\
  120 & 0.13 & 1.00 & 0.740 & 1.44 & 0.30 & 0.04 & 0.78 & 1.07 & 1.50 \\
  \multicolumn{10}{c}{Canup (2004)} \\
  120& 0.13 & 1.00 & 0.730 & 1.59 & 0.30 & 0.05 & 0.77 & 1.16 & 1.28 \\
  120& 0.13 & 1.00 & 0.740 & 1.54 & 0.31 & 0.04 & 0.80 & 1.19 & 1.44 \\
\enddata
\tablecomments{ $N$ -- number of particles, $\gamma$ -- mass ratio, $v$
  -- impact velocity in units of mutual escape velocity, b -- impact
  parameter, $M_D$ -- disk mass, $M_L$ -- Lunar mass, $L_D$ -- disk
  angular momentum, $L_{EM}$ -- angular momentum of Earth-Moon system,
  $M_{Fe}$ -- mass of iron in disk, $M_{imp}$ -- mass of disk
  originating from impactor, $L_F$ -- final system angular momentum,
  and $M_M$ -- estimated mass of satellite forming from disk.}
\label{ref:MFI}
\end{deluxetable}

\clearpage

\clearpage

\clearpage

\clearpage

\clearpage
\bibliographystyle{apj_fixed}

\end{document}